\documentclass{article}
\usepackage[utf8]{inputenc}
\usepackage{appendix}
\usepackage[]{footmisc}
\usepackage{amsmath}
\usepackage{cite}
\usepackage{textcomp}
\usepackage{hyperref}

\title{Matrix-based Induced Metric with Non Commutative Gauge Fields}
\author{Leandro de Paula\thanks{Email:\href{mailto:leandroifusp@yahoo.com.br}{leandroifusp@yahoo.com.br}}\\
University of Maryland }
\date{\today}

\begin{document}
\maketitle

\begin{abstract}
    This work presents a matrix formulation for the induced metric on a brane, where non-commutative gauge fields play a central role. By starting from a general embedding of a p-brane in a higher-dimensional bulk, we derive the induced metric using commutator relations and extend the framework by incorporating gauge fields through a minimal coupling prescription. This approach not only draws an analogy with the effective acoustic metric in condensed matter systems but also provides insight into the emergence of light quanta as low-energy collective excitations of the quantum vacuum. In a simplified setting with a free Hamiltonian, we recover standard results in the appropriate limits while also uncovering new regimes characterized by a non-trivial coupling between the emergent gauge fields and the underlying matrix degrees of freedom. Our findings suggest potential pathways for connecting aspects of string theory matrix models with phenomenological features of emergent gravity.
\end{abstract}

\section{Introduction}

Braneworld scenarios provide a fertile framework in modern theoretical physics, proposing that our observable universe is a four-dimensional hypersurface (brane) embedded in a higher-dimensional space (bulk). This framework not only challenges conventional notions of spacetime but also offers elegant solutions to long-standing problems in physics, such as the hierarchy problem and the cosmological constant puzzle \cite{brax2004, koyama2007, maartens2010}. 

The origin of modern braneworld scenarios can be traced back to M-theory, which is proposed to unify all five consistent string theories into a single theoretical framework. M-theory extends string theory by introducing an 11-dimensional bulk, within which our universe is confined to a 10-dimensional brane \cite{horava1995}. In this framework, our observable four-dimensional spacetime is envisioned as a slice of the higher-dimensional bulk, with six spatial dimensions compactified into Calabi-Yau manifolds. The remaining extra dimension is structured as an orbifold \(S^1/Z_2\), creating a geometric separation between two distinct branes that serve as boundaries of the higher-dimensional space. One of these branes hosts the familiar four-dimensional Standard Model, while the other remains elusive and may harbour new physics. 

The Randall-Sundrum (RS) model, for instance, is one of the scenarios that capture certain aspects of M-theory by proposing a 5-dimensional warped-geometry spacetime, where our universe is represented as a four-dimensional slice located at its boundary. Specifically, the RS1 model features two branes bounding the extra dimension, whereas the RS2 model extends the extra dimension infinitely, placing one brane at infinity \cite{randall1999, sundrum1999}.  

The notion of a brane can also be found in condensed matter systems, particularly, in systems where topological defects form \cite{Volovik2003}. These defects are regions in which the order parameter (which characterizes the phase of the system) behaves differently from that in the surrounding bulk. An intriguing aspect of these defects is their lower-dimensional nature, which shares conceptual similarities with braneworld scenarios in theoretical physics. In many materials, defects such as domain walls, vortices, and monopoles separate regions with different ordering or symmetry, much like how branes delineate regions in higher-dimensional theories. These defects can trap localized modes of excitation and significantly influence the system's dynamics. For example, vortices in superfluid helium phase A (\(^3\)He-A) can give rise to the emergency of collective excitations and quasi-particules that mimic the gauge fields and fundamental particles of the Standard Model.  

This paper is part of a broader research program that leverages insights from analogs in condensend matter systems as well as from more fundamental spacetime theories to aid in the development of a theory of everything \cite{Volovik2008}.  This cross-disciplinary approach not only deepens our understanding of complex condensed matter phenomena but also provides a tangible way to explore and test concepts that are central to high-energy physics. In particular, here, we aim to connect the concepts of a brane induced metric with the so-called effective acoustic metric. Throughout this paper one uses natural units \(\hbar=c=1\), unless stated otherwise.

\section{The operator-based metric }

A p-brane embedded within N-dimensional spacetime represents an extended object in \(p+1\) dimensions. Its intrinsic geometry is described by the induced metric, which is derived from the background spacetime metric via a pullback operation:

\begin{equation}
    ds^2=g_{\mu \nu}(X) d X^\mu d X^\nu=g_{\mu \nu}(X) \frac{\partial X^{\mu}}{\partial \xi^\alpha} \frac{\partial X^\nu}{\partial \xi^{\beta}}d\xi^\alpha d\xi^\beta, \label{eq1}
\end{equation}
where
\renewcommand{\labelitemi}{-}

\begin{itemize}
    \item \(X^\mu\ (\mu = 0, 1, 2, ..., N) \) are coordinates in the background spacetime;
    \item  \(\xi^\alpha\ (\alpha = 0, 1, 2, p < N) \) are coordinates on the brane’s worldvolume;
    \item \( X(\xi) \) are the embedding functions describing the position of the p-brane within the N-dimensional spacetime;
    \item \( g_{\mu\nu}(X) \) represents the metric of the background spacetime.
\end{itemize}

In flat spacetime \(g_{\mu \nu}(X)\) is just the Minkowski metric \(\eta_{\mu\nu} = \text{diag}(-1, 1, 1, 1,...)\) and the induced metric is written as:

\begin{equation}
    g_{\alpha\beta} = \eta_{\mu\nu} \partial_{\alpha} X^{\mu} \partial_{\beta} X^{\nu}. \label{eq2}
\end{equation}

Taking advantage of the gauge freedom on the brane, one can locally make \(X^\mu=\delta^\mu_\alpha\xi^\alpha+a^\mu\), where the term \( \delta^\mu_\alpha \xi^\alpha \) aligns the brane's intrinsic coordinates (\( \xi^\alpha \)) with the first \( p+1 \) dimensions of the higher-dimensional space and the \( a^\mu \) component indicates a constant shift, ensuring the brane's position isn't necessarily tied to the origin. This is a trivial embedding and thus \(\partial_{\alpha} X^\mu=\delta^\mu_\alpha\). It represents a local flat brane embedded in the ambient spacetime with

\begin{equation}
    g_{\alpha\beta} = \eta_{\mu \nu} \delta^\mu_\alpha \delta^\nu_\beta=\eta_{\alpha\beta}, \label{eq3}
\end{equation}
where the induced metric also becomes a Minkowski metric. With this embedding, one sees that the induced metric inherits bulk's properties.

However, let us keep the embedding more general in eq. \eqref{eq2}. We also treat the embedding functions as dynamical variables so that they become operators in the Hilbert space. A displacement in the worldvolume coordinates, \(\xi \rightarrow \xi' = \xi + \lambda\), implies a corresponding displacement in the spacetime coordinates, represented as

\begin{equation}
X^\mu(\xi + \lambda) = e^{i \lambda^\alpha P_\alpha} X^\mu(\xi) e^{-i \lambda^\alpha P_\alpha} = X^\mu(\xi) + i \lambda^\alpha [P_\alpha, X^\mu(\xi)] + \mathcal{O}(|\lambda|^2), \label{eq4}
\end{equation}
where \(\lambda\) denotes an infinitesimal parameter, and \(P_\alpha\) represents the generators of translations in the worldvolume directions. By comparing equation \eqref{eq4} with the Taylor expansion

\begin{equation}
X^\mu(\xi + \lambda) = X^\mu(\xi) + \lambda^\alpha \partial_\alpha X^\mu(\xi) + \mathcal{O}(|\lambda|^2), \label{eq5}
\end{equation}
we obtain the relation:

\begin{equation}
i \partial_\alpha X^\mu = [X^\mu, P_\alpha]. \label{eq6}
\end{equation}
Replacing the previous eq. \eqref{eq6} into eq. \eqref{eq2}, we get:

\begin{equation}
g_{\alpha\beta} = -[X^{\mu}, P_{\alpha}][X_{\mu}, P_{\beta}]. \label{eq7}
\end{equation}

\subsection{Analogy with effective acoustic metric}

Symmetries, conservation laws, and analogies are powerful tools for probing theoretical models, and condensed matter analogs can help us understand spacetime. In particular, certain aspects of condensed matter physics have given rise to the field of ``analogue gravity'', which uses laboratory systems to investigate features of curved spacetime \cite{Barcelo2011}. According to this perspective, gravity and spacetime are emergent phenomena, and quantum hydrodynamics may offer valuable insights into a quantum theory of gravity \cite{Volovik2008}. For instance, the behavior of sound waves in a moving fluid provides a simple analogue model that mimics the propagation of light in curved spacetime. These models rely on a mathematical construct called the effective acoustic metric, which serves as an analogue for the metric of curved spacetime.

A fluid moving relative to the laboratory generates sound waves that propagate relative to the fluid. In the laboratory frame, the velocity of a sound ray traveling in the direction of the unit vector \(\vec{n}\) (where \(\vec{n}^2=1\)) is given by

\begin{equation}
    \frac{d\vec{x}}{dt} = c_s \vec{n} + \vec{v} \label{eq8}
\end{equation}
where \(c_s\) is the speed of sound relative to the fluid, and \(\vec{v} = \vec{v}(t,\vec{r})\) is the speed of the
fluid relative to the laboratory at the instant \(t\) and position \(\vec{r}\). By handling eq. \eqref{eq8}, one obtains the relation

\begin{equation}
    (d\vec{x} - \vec{v}\,dt)^2 - c_s^2\,dt^2 = 0, \label{eq9}
\end{equation}
which expresses the null condition for sound propagation (i.e., \(ds^2 = 0\)). This can be recast as

\begin{equation}
ds^2 = -\left(1 - \frac{v^2}{c_s^2}\right)(c_s\,dt)^2 - 2\,\frac{\vec{v}}{c_s}\cdot d\vec{x}\,(c_s\,dt) - d\vec{x}\cdot d\vec{x}, \label{eq10}
\end{equation}
where \(v^2 = \vec{v}\cdot\vec{v}\).

From eq. \eqref{eq10}, an effective metric for sound propagation in the fluid is defined as

\begin{equation}
g_{00} = -1 + \frac{v^2}{c_s^2},\quad g_{0i} = -\frac{v_i}{c_s},\quad g_{ij} = \delta_{ij}. \label{eq11}
\end{equation}

This effective acoustic metric shows how the motion of the fluid influences the propagation of sound, analogous to how a curved spacetime metric governs the motion of light. A null line element admits the omission of a multiplicative prefactor. Thus no conformal
factor has been explicit in \eqref{eq11}.

Imposing the commutation relations \([X^i, P_j] = i \delta^i_j\) and \([X^0, P_\alpha] = i \delta^0_\alpha\), the components of the metric operator in eq. \eqref{eq7} reduce to:
\begin{subequations}
\begin{align}
g_{00} &= -[X^\mu, P_0][X_\mu, P_0] = -I - [X^i, H][X_i, H], \label{eq12a}\\
g_{0j} &= -[X^\mu, P_0][X_\mu, P_j] = i\,[X_j, H], \label{eq12b}\\
g_{ij} &= -[X^\mu, P_i][X_\mu, P_j] = \delta_{ij}\, I. \label{eq12c}
\end{align} \label{eq12}
\end{subequations}

From eq. \eqref{eq6}, one can think of those imposed commutation relations as a gauge fixing with \(\partial_j X^i = \delta^i_j\) and \(\partial_{\alpha} X^0 = \delta^0_{\alpha}\). The only remaining degrees of freedom left is the temporal evolution of \(X^i\) given as \(i \partial_t X^i = [X^i, H]\), where \(P_0=H\) is a Hamiltonian. 

Note that a velocity operator is generally defined in terms of a Hamiltonian operator as \(iV_i = [X_i, H]\). Also, let \(\bar{v}_i(\xi)\) denote the mean value of this velocity operator in a state \(|\psi\rangle\), i.e., \(\bar{v}_i = \langle \psi | V_i | \psi \rangle\). Inserting these observations into eqs. \eqref{eq12}, one obtains: 

\begin{equation}
\bar{g}_{00} = -1 + \frac{\bar{v}_i \bar{v}_i}{c^2}, \quad \bar{g}_{0j} = -\frac{\bar{v}_j}{c}, \quad \bar{g}_{ij} = \delta_{ij}, \label{eq13}
\end{equation}
where the velocity of light (\(c\)) has been reinserted.

Let us consider the effective geometry \eqref{eq11} as an analogue model for the mean value of the quantum geometry \eqref{eq13}, with \(c_s\) and \(v_i(t,\vec{r})\) playing the roles of \(c\) and \(\bar{v}_i(\xi)\), respectively. It is well known that the effective geometry emerges from the dynamics of a fluid. Here, by identifying \(c \to c_s\) and \(\bar{v}_i(\xi) \to v_i(t,\vec{r})\), the effective geometry also emerges from the quantum geometry \cite{Volovik1979}. In this case, it is reasonable to interpret the Hamiltonian as describing the degrees of freedom that constitute the quantum fluid. On the other hand, because \eqref{eq12} can be derived solely from considerations of spacetime symmetries, its meaning is more fundamental. In other words, the Hamiltonian describes the degrees of freedom of the quantum vacuum whose dynamics give rise to the metric field.

\subsection{Application to free hamiltonian}

The metric operator equations have been expressed in terms of a hamiltonian operator in \eqref{eq12}. One can consider the simplest case of a free hamiltonian given by

\begin{equation}
H = l_0 P_i^2 / 2 + E_0, \label{eq14}
\end{equation}
where \(l_0\) is a characteristic length and \(E_0\) is an internal energy contribution that commute with \(X^i\) although \([X^0,E_0]=i\). The commutation relations then yield:

\begin{equation}
[X^i, H] = i l_0 P^i. \label{eq15}
\end{equation}

Substituting eq. \eqref{eq15} into the metric operator \eqref{eq12}, we obtain:

\begin{equation}
g_{00} = -I + l_0^2 P^i P_i, \quad g_{0j} = -l_0 P_j, \quad g_{ij} = \delta_{ij} I. \label{eq16}
\end{equation}
If \(K^i\) and \(|K^i\rangle\) are the eigenvalues and eigenvectors of the momentum operator \(P^i\), then \(P^i|K^i\rangle= K^i|K^i\rangle\). The contravariant components of eigenvalues in \eqref{eq16} are given by
\begin{equation}
g^{00} = -1,\quad g^{0i} = -l_0 K^i,\quad g^{ij} = \delta^{ij}-l^2_0 K^iK^j \label{eq17}
\end{equation}

The model given by \eqref{eq14} possess a ground state of energy. Following our analogy with condensed matter, “light quanta”  may emerge as low-energy collective excitations propagating over this vacuum. The energy spectrum of these low-energy collective excitations (light quanta) is given by:

\begin{equation}
g^{\alpha\beta} p_\alpha p_\beta = 0, \label{eq18}
\end{equation}
where \(p_\alpha = (-E, p_1, p_2, p_3,...)\) represents the four-momentum of the quanta. This equation demonstrates that the spectrum is also independent of any multiplicative prefactor (conformal factor). From \eqref{eq18} and \eqref{eq17}, one obtain:

\begin{equation}
-E^2+2l_0K^ip_iE+p^2-l^2_0 (K^ip_i)^2 = 0,\label{eq19}
\end{equation}
where \(p^2=\|p\|^2=\delta^{ij}p_ip_j\). By handling eq. \eqref{eq19}, we get:

\begin{equation}
E = l_0 K^i p_i + p. \label{eq20}
\end{equation}

This energy spectrum differs from the standard spectrum due to the presence of an additional term involving the characteristic length \((l_0)\), which breaks Lorentz invariance \cite{Jacobson2006}. By setting \(l_0 = 0\), one recovers the standard spectrum for massless relativistic particles, \(E=p\). In particular, let us consider the case where \(p=p_1\) so that \(E = (l_0K^1 + 1)p_1\). If \(l_0\neq 0\), two distinct regimes emerge. For \(|l_0K^1|\ll 1\), the standard relation \(E=p\) holds once again. However, in the regime where \(l_0K^1\gg 1\), the “light quanta” become more strongly coupled to the fundamental degrees of freedom, leading to \(E=l_0K^1p_1\). By employing dimensional analysis, one can estimate the energy associated with \(K^1\) to be of the order of thermal energy, i.e., \(l_0(K^1)^2 \approx kT\), or equivalently, \(|l_0K^1|\approx \sqrt{l_0kT}\). Assuming \(l^{-1}_0 \approx 10^{19} \ \text{GeV}\), we find that for the current temperature of the Universe, \(kT\approx\  10^{-4}\ \text{eV}\), yelding \(|l_0K^1|\approx 10^{-16}\ (\ll 1)\). On the other hand, in the early Universe, where \(kT\approx 10^{19}\  \text{GeV}\), we obtain \(l_0 K^1\approx 1\). Finally, the regime \(|l_0K^1|\gg 1\) corresponds to the transplanckian domain.

\section{The non-commutative gauge fields}

In the minimal coupling prescription, gauge fields are introduced by modifying the momentum and the derivative to ensure gauge invariance. The generalized momentum and covariant derivative incorporating gauge fields are given by:

\begin{equation}
\Pi_{\alpha} = P_{\alpha} - q A_{\alpha}(\xi) \label{eq21}
\end{equation}
and
\begin{equation}
    D_\alpha X^\mu=\partial_\alpha X^\mu+iq[X^\mu,A_\alpha], \label{eq22}
\end{equation}
where:
\begin{itemize}
    \item \(q\) is the charge associated with the gauge field;
    \item \(A_{\alpha}(\xi)\) is the gauge potential that depends on the worldvolume coordinates.
\end{itemize}
By inserting \eqref{eq21} and \eqref{eq22} into eq. \eqref{eq6}, we obtain:

\begin{equation}
i D_\alpha X^\mu = [X^\mu, \Pi_\alpha]. \label{eq23}
\end{equation}
Therefore, equation \eqref{eq2} or \eqref{eq7} is modified to include gauge fields:

\begin{equation}
g_{\alpha \beta} = - [X^{\mu}, \Pi_{\alpha}]\star [X_{\mu}, \Pi_{\beta}]. \label{eq24}
\end{equation}
Note that the star product \(\star\) has been introduced. From now on, noncommutative geometry will be considered, as it is well-motivated in string theory, where it naturally arises under certain conditions.

The star commutators in eq. \eqref{eq24} is expanded and calculated as below:

\begin{equation}
[X^{\mu}, \Pi_{\alpha}]_\star = [X^{\mu}, P_{\alpha}]_\star - q [X^{\mu}, A_{\alpha}]_\star.\label{eq25}
\end{equation}
The first commutator on the right hand side of eq. \eqref{eq25} is further expanded (see Appendix):
 
\begin{equation}
[X^\mu, P_\alpha]_\star = [X^\mu, P_\alpha] + (i/2) \theta^{\gamma\lambda} \{\partial_\gamma X^\mu, \partial_\lambda P_\alpha\}, \label{eq26}
\end{equation}
where \(\theta^{\gamma\lambda}\) is the non commutative parameter. The second term in \eqref{eq26} is zero because the derivative of the momentum in a free Hamiltonian vanishes. The first term has already been explored in the previous sections. Recalling it once again:

\begin{equation}
    [X^i, P_j] = i \delta^i_j,\quad [X^0, P_\alpha] = i \delta^0_\alpha, \quad \text{and}\quad [X^i, H] = i l_0 P^i. \label{eq27}
\end{equation}
For simplicity, we consider now \(P^i = 0\) , i.e., \(qA_{\alpha}>>P_{\alpha}\). This is also the regime where the gauge fields \(A_\alpha(\xi)\) interact weakly with the substrate. Thus, eq. \eqref{eq6} becomes

\begin{equation}
i \partial_\alpha X^\mu = [X^\mu, P_\alpha]_{\star} = i\delta^\mu_\alpha. \label{eq28}
\end{equation}

The second commutator in eq. \eqref{eq25} is also expanded in the form:

\begin{equation}
[X^\mu, A_\alpha]_\star = [X^\mu, A_\alpha] + (i/2) \theta^{\gamma \lambda} \{\partial_\gamma X^\mu, \partial_\lambda A_\alpha\},\label{eq29}
\end{equation}
The first term on the second-hand side is the usual commutator, and it vanishes. \(X^\mu\) and \(A_\alpha\) are independent degrees of freedom at zeroth order of non commutativity. From \eqref{eq28} and \eqref{eq29}:

\begin{equation}
[X^\mu, A_\alpha]_\star = i \theta^{\gamma \lambda} \delta_\gamma^{\mu} \partial_\lambda A_\alpha.\label{eq30}
\end{equation}
For brevity, we use the contraction \(\theta^{\mu \lambda} = \theta^{\gamma \lambda} \delta_\gamma^{\mu}\), assigning \(\theta^{a \lambda} = 0\), where \(a\) labels the extra-dimensional indices. 

Expanding for the electromagnetic, weak, and strong gauge fields \(A\), \(W\), and \(G\), with their corresponding charges \(e\), \(q\), and \(g\), respectively:

\begin{equation}
\begin{aligned}
    g_{\alpha \beta} &=
        \eta_{\mu \nu} \left( 
            -i \delta^\mu_\alpha 
            + ie \theta^{\mu \rho} \partial_\rho A_\alpha 
            + iq \theta^{\mu \rho} \partial_\rho W_\alpha^i T^i
            + ig \theta^{\mu \rho} \partial_\rho G_\alpha^i T^i 
        \right) \\
    &\quad\times \left( 
            i \delta^\nu_\beta 
            - ie \theta^{\nu \lambda} \partial_\lambda A_\beta 
            - iq \theta^{\nu \lambda} \partial_\lambda W_\beta^j T^j 
            - ig \theta^{\nu \lambda} \partial_\lambda G_\beta^j T^j 
        \right),
\label{eq31}
\end{aligned}
\end{equation}
where \(\eta_{\mu\nu}\) has been reinserted. Expanding the terms, we obtain:

\begin{subequations}\label{eq32}
\begin{align}
    g_{\alpha \beta} &= \eta_{\alpha\beta} + \eta_{\mu \nu} \Big( \nonumber \\
    &\quad 
        - e \delta^\mu_\alpha \theta^{\nu \lambda} \partial_\lambda A_\beta 
        - e \theta^{\mu \rho} \partial_\rho A_\alpha \delta^\nu_\beta 
        + e^2 \theta^{\mu \rho} \theta^{\nu \lambda} \partial_\rho A_\alpha \partial_\lambda A_\beta \label{eq32a} \\
    &\quad 
        - q \delta^\mu_\alpha \theta^{\nu \lambda} \partial_\lambda W_\beta^j T^j 
        - q \theta^{\mu \rho} \partial_\rho W_\alpha^i T^i \delta^\nu_\beta 
        + q^2 \theta^{\mu \rho} \theta^{\nu \lambda} \partial_\rho W_\alpha^i \partial_\lambda W_\beta^j T^i T^j \label{eq32b} \\
    &\quad 
        - g \delta^\mu_\alpha \theta^{\nu \lambda} \partial_\lambda G_\beta^j T^j 
        - g \theta^{\mu \rho} \partial_\rho G_\alpha^i T^i \delta^\nu_\beta 
        + g^2 \theta^{\mu \rho} \theta^{\nu \lambda} \partial_\rho G_\alpha^i \partial_\lambda G_\beta^j T^i T^j \label{eq32c} \\
    &\quad 
        + e q \theta^{\mu \rho} \theta^{\nu \lambda} \partial_\rho A_\alpha \partial_\lambda W_\beta^j T^j 
        + e q \theta^{\mu \rho} \theta^{\nu \lambda} \partial_\lambda A_\beta \partial_\rho W_\alpha^i T^i \label{eq32d} \\
    &\quad 
        + eg \theta^{\mu \rho} \theta^{\nu \lambda} \partial_\rho A_\alpha \partial_\lambda G_\beta^j T^j 
        + eg \theta^{\mu \rho} \theta^{\nu \lambda} \partial_\lambda A_\beta \partial_\rho G_\alpha^i T^i \label{eq32e} \\
    &\quad 
        + qg \theta^{\mu \rho} \theta^{\nu \lambda} \partial_\rho W_\alpha^i \partial_\lambda G_\beta^j T^i T^j 
        + qg \theta^{\mu \rho} \theta^{\nu \lambda} \partial_\lambda W_\beta^j \partial_\rho G_\alpha^i T^iT^j \Big). \label{eq32f}
\end{align}
\end{subequations}
We can further simplify the expression in parentheses in eq. \eqref{eq32} by defining it as \(\Lambda_{\alpha\beta}^{\mu\nu}\):

\begin{equation}
g_{\alpha \beta} =  \eta_{\alpha\beta} +\eta_{\mu \nu}\Lambda^{\mu \nu}_{\alpha\beta}. \label{eq33}
\end{equation}
By taking the expectation value on both sides, one can get:

\begin{equation}
\delta g_{\alpha \beta} =  \eta_{\mu \nu}\langle \Lambda^{\mu \nu}_{\alpha\beta}\rangle, \label{eq34}
\end{equation}
where \(\delta g_{\alpha \beta} = \langle g_{\alpha \beta} \rangle - \eta_{\alpha \beta}\) represents perturbations in the 4D part of the bulk metric due to the dynamics of noncommutative gauge fields. Gauge fields are treated as perturbations of the substrate, analogous to how phonons are seen as perturbations of an atomic condensate.

\section{Discussion}

The construction presented in this manuscript leverage insights from both string theory and condensed matter physics. Condensed matter systems can be described using effective quantum field theory without much reference to the true degrees of freedom - namely, the underlying collection of atoms that compose the system. Although there is a Hamiltonian involving the kinematics and interactions of the fundamental atoms, many insights can be extracted from the effective theory. 

As we derived, the brane-induced metric bears an analogy to the acoustic metric in the appropriate gauge. If the embedding functions are promoted to operators, the induced metric also becomes an operator in the Hilbert space. One can take this further by considering the embedding functions as matrix-valued quantities, similar to BFSS-type models \cite{BFSSmatrix1997}. The BFSS matrix model is conjectured to be equivalent to M-theory in the infinite momentum frame (IMF), where Galilean invariance holds. Our brane-induced metric is designed to match the galilean form of the acoustic metric. Consequently, the resulting matrix-valued induced metric can be interpreted as a structure valid in the IMF, where the infinite momentum is associated with the 11th dimension of M-theory or the 5th dimension in Randall-Sundrum models.

We have not specified our fundamental degrees of freedom. In the BFSS model, the fundamental degrees of freedom are D0-branes - ``point particles'' - interacting according to the Hamiltonian (excluding supersymmetric terms):

\begin{equation}
H = \text{Tr} \left( \frac{P_i^2}{2} - \frac{1}{4} [X^i, X^j]^2\right) \label{eq35}
\end{equation}
Compared to our ansatz in Eq. \eqref{eq14}, this Hamiltonian is both more specific and broader. It is more specific because our total momentum in Eq. \eqref{eq14} can be expanded as a collection of D0-brane momenta: \(P_i^{\text{total}} = \text{Tr}(P_i^2)\). It is broader because the ground state energy in Eq. \eqref{eq14}, \(E_0\), also includes interacting terms between coordinates, as seen in Eq. \eqref{eq35}. One way to recover the ground state \(E_0\) is to restrict the interaction term using the non-commutative parameter as \([X^i, X^j] = i\theta^{ij}\), so that \(E_0 = \text{Tr}(\theta^2 / 4)\), where \(\theta = \theta^{ij}\theta_{ij}\).  

The acoustic-type metric can emerge from the brane-induced metric, where the collective dynamics of D0-branes moving in the infinite momentum frame induce effective spacetime geometries that govern the propagation of excitations in the brane system. These excitations can be treated as acoustic waves, described by a metric analogous to those in fluid dynamics or fluid-gravity analogs. In this context, the minimal coupling prescription for gauge fields introduces perturbations in the momentum of the brane, just as perturbations occur in the density or velocity of the acoustic metric. Similarly, \(\delta g_{\alpha \beta} = \langle g_{\alpha \beta} \rangle - \eta_{\alpha \beta}\) represents perturbations in the 4D part of the bulk metric due to the dynamics of noncommutative gauge fields. Gauge fields are viewed as perturbations of the substrate, analogous to how phonons are seen as perturbations of an atomic condensate.

If one considers the braneworld as a collection of fundamental degrees of freedom in the infinite momentum frame (IMF), the Galilean regime used in our acoustic metric and Hamiltonian makes sense. Our task is to align these with more fundamental physics tools, such as the brane-induced metric and the BFSS Hamiltonian.

After gauge fixing, Eqs. \eqref{eq12} are derived. The remaining degrees of freedom are those that are time-dependent, governed by the equation

\begin{equation}
i \partial_t X^i = [X^i, H].\label{eq36}
\end{equation}
This is also the Heisenberg evolution equation for the degrees of freedom represented by the coordinates \(X^i\), which correspond to the D0-branes in matrix theories. Although the dependence of \(X^i\) on other parameters is pure gauge and does not involve dynamics, these parameters still serve to localize the D0-branes.

At very small scales, the Minkowski metric provides a good approximation in the absence of strong gravitational fields; however, when strong gravitational effects are present, the curvature of spacetime cannot be neglected. The minimal coupling prescription for gravity can be applied in Eq. \eqref{eq24} by replacing \(-\eta_{\mu \nu}\) with \(g_{\mu \nu}(X)\). The general form of the metric matrix then becomes:

\newcommand{\mystrut}{\rule[-2ex]{0pt}{5ex}}
\begin{equation}
\fbox{\makebox[2in][c]{\mystrut \(g_{\alpha \beta} = g_{\mu \nu} [X^{\mu}, \Pi_{\alpha}] [X^{\nu}, \Pi_{\beta}]\)}}.\label{eq37}
\end{equation}
Here, the minus sign is absorbed into the metric, and the star product is omitted since the matrix coordinates are already noncommutative.

\begin{appendices}

\section{Appendix Section A: Moyal star product}

The Moyal star product is a way of introducing noncommutativity in the product of functions. For two functions \( f(x) \) and \( g(x) \) on a space with coordinates \( x \), the product is defined as

\begin{equation}
(f \star g)(x) = f(x) \exp\left( \frac{i}{2} \theta^{\mu \nu} \overleftarrow{\partial_\mu} \, \overrightarrow{\partial_\nu} \right) g(x), \label{apeq1}
\end{equation}
where \( \theta^{\mu \nu} \) is a constant antisymmetric matrix that determines the noncommutativity, and \( \overleftarrow{\partial_\mu} \) and \( \overrightarrow{\partial_\nu} \) represent derivatives acting to the left and right, respectively. 

Expanding the exponential operator, the star product up to the second order can be written as:

\begin{equation}
    (f \star g)(x) = f(x) g(x) + \frac{i}{2} \theta^{\mu \nu} (\partial_\mu f)(\partial_\nu g) - \frac{1}{8} \theta^{\mu \nu} \theta^{\alpha \beta} (\partial_\mu \partial_\alpha f)(\partial_\nu \partial_\beta g) + \dots \label{apeq2}
\end{equation}

If we take \( f = X^\mu \) and \( g = A_\alpha \), where \( X^\mu \) is the embedding functions and \( A_\alpha \) are the component of a gauge field in function of the world-volume coordinates, we obtain:

\begin{equation}
    (X^\mu \star A_\nu)(\xi) = X^\mu A_\nu + \frac{i}{2} \theta^{\alpha\beta} \partial_\alpha X^\mu\partial_\beta A_\nu, \label{apeq3}
\end{equation}
 
The star commutator is given by:

\begin{align}
[X^\mu, A_\nu]_\star & = 
 (X^\mu \star A_\nu)-(A_\nu \star X^\mu) \nonumber\\
 & =  X^\mu A_\nu - A_\nu X^\mu + \frac{i}{2} \theta^{\alpha\beta} \partial_\alpha X^\mu\partial_\beta A_\nu-\frac{i}{2} \theta^{\beta\alpha} \partial_\beta A_\nu \partial_\alpha X^\mu \nonumber\\
 & = [X^\mu, A_\alpha] + (i/2) \theta^{\gamma \lambda} \{\partial_\gamma X^\mu, \partial_\lambda A_\alpha\}, \label{apeq4}
\end{align}
where we used \([\ ,\ ]\) and \(\{\ ,\ \}\) as commutator and anticommutator brackets, respectively, and \(\theta^{\alpha\beta}=-\theta^{\beta\alpha}\),
as in eq. \eqref{eq29}.

\end{appendices}

\end{document}